# Intelligence Complements from the Built Environment: A review of Smart Building Technologies for Cognitively Declined Occupants

Saeid Alimoradi [a] and Xinghua Gao [b*]

[a] Ph.D. Student, Myers-Lawson School of Construction, Virginia Polytechnic Institute and State University, 1345 Perry Street, Blacksburg, VA, USA, 24061; e-mail: salimora@vt.edu

[b*] Assistant Professor, Myers-Lawson School of Construction, Virginia Polytechnic Institute and State University, 1345 Perry Street, Blacksburg, VA, USA, 24061; e-mail: xinghua@vt.edu

**Abstract**

Traditionally, caregivers, whether formal or informal, have taken the responsibility of providing assistance and care to patients with cognitive decline. Usually, both the caregiver and the patient are subjected to financial and emotional burdens, which impact the patient's life quality. To overcome this issue, Ambient Assistive Living (AAL) technologies have been adopted to replace the caregivers and complement patients' lack of intelligence. Technologies such as Internet of Things (IoT) and Artificial Intelligence (AI) have enabled intelligent ubiquitous learning for smart buildings to monitor the cognitively declined occupants and provide in-home assistive services and solutions. This paper aims to summarize and evaluate the intelligence complements provided by smart buildings that can increase the cognitively declined occupants' quality of life and autonomy. Through a systematic literature review, the authors find that most of the existing contributions are towards identifying the occupants' behavior, and thus, to determine corresponding assistive services and solutions. Five key research gaps are identified, including the lack of adequate adoption of technological interventions to fully support the occupants' autonomy and independence. The authors also propose a conceptual framework to highlight the research gaps in smart building applications for cognitively declined occupants and to map the future research directions.

## 1 Introduction

The common difficulties faced by people diagnosed with cognitive decline include problem-solving, memory, comprehension, and attention [1]. Because of such difficulties, people with cognitive decline tend to act incoherently and perform tasks more erroneously than healthy people [2,3]. Their capacity to perform simple tasks and everyday instrumental activities are decreased, and thus, cannot live independently at home [2,3]. Activities such as cooking, getting dressed, grooming, bathing, or housework, in general, are considered complicated for the patients to initiate or complete [4,5]. To successfully perform activities of daily living (ADLs), the patients require institutionalized or hospitalized care or to be taken care of by family members or caregivers at home [6]. Because of the forgetfulness associated with cognitive decline, caregivers normally have to deliver numerous reminders to patients. In more advanced stages of cognitive decline, caregivers, formal or informal, are required to intervene and carry out a portion of the activities as compensation for the lack of intelligence in performing ADLs.

Providing sufficient recourses to train professional caregivers, and long-term hospitalizations or institutionalizations are a major societal and economic burden, which affects both public finances and the provision of healthcare services [7]. Furthermore, many patients prefer to stay at home [8,9]. Informal caregivers struggle with psychological burdens such as emotional stress, anxiety, and depression since they perform heavy caregiving tasks without proper compensations. The dependency on others to perform simple everyday activities draws an emotional pressure on the patients as well [7,10,11]. As a result, the patients' life quality and the provided care drop significantly with care home admissions [12,13].

Technology has been beneficial in improving the patients' life quality through automation of caregiving responsibilities [14]. Provision of support as an intelligence compensation promotes cognitively declined occupants' independence, diminishes the burden on public finances, relieves the struggles caregivers experience, and enhances and optimizes quality of care [1,12,15].

Recent developments in the fields of Artificial Intelligence (AI), Internet of Things (IoT), robotics, and human-computer interaction, have enabled new possibilities for in-home care of people with cognitive decline without the help of caregivers [7,12]. The application of such developments can be analyzed within the larger domain of Cyber-Physical Systems (CPS). CPS is an incorporation of processes into the physical environment through robotics and human-computer interactions with computational (AI) and communicational (IoT) components [16–18]. Delivering therapeutic or assistive interventions into the built environments through controlling the planted actuators and effectors is an application of CPS in the smart building domain to provide intelligent compensations to the cognitively declined occupants.

CPS and IoT-enabled smart buildings have been a topic of interest. However, there is a need to understand what smart buildings can offer to people with cognitive decline and how the ubiquitous intelligence of the built environment can complement the decrease in occupants' intelligence. In this study, the authors provide insights for researchers and industry professionals into the supports of the smart building for the cognitively declined occupants. The study further evaluates the previous adoption of CPS, IoT enabled sensing, and learning capabilities supporting independence of occupants with cognitive decline. This research explores and explains the current research developments, identifies research gaps, and provides future research directions. The remainder of the study is as follows. Section 2 defines the scope of the research and explains the methodology of the study to find academic articles in the scope. Section 3 overviews the current contributions that smart buildings can offer to the cognitively declined occupants to support their independency. Section 4 discusses the research gaps and limitations and provides suggestions and directions for the future research. Section 5 proposes a conceptual CPS framework for automation of caregiving responsibilities and complementing the lack of intelligence in occupants. Section 6 presents the conclusions.

## 2 Methodology

This scoping review was conducted to identify, evaluate, and summarize the contributions in technological solutions enhancing the QoL for people with cognitive impairment and improving their independence. The review process, which was inspired by [19], included four major steps (Figure 1): 1) definition of the research scope and identification of search terms to be used in search databases, 2) collection of the documents relevant to the defined scope, 3) analysis the content and identifying the advancements and developments made hitherto in various components of assistive technologies utilized to help people with cognitive impairment to live independently (Section 3), 4) evaluation of the identified research gaps, challenge, and limitations, and suggestions and strategies for the future research (Section 4), and introduction of an ideal framework for a CPS-enabled smart home based on the current and existing contributions and identified gaps (Section 5).

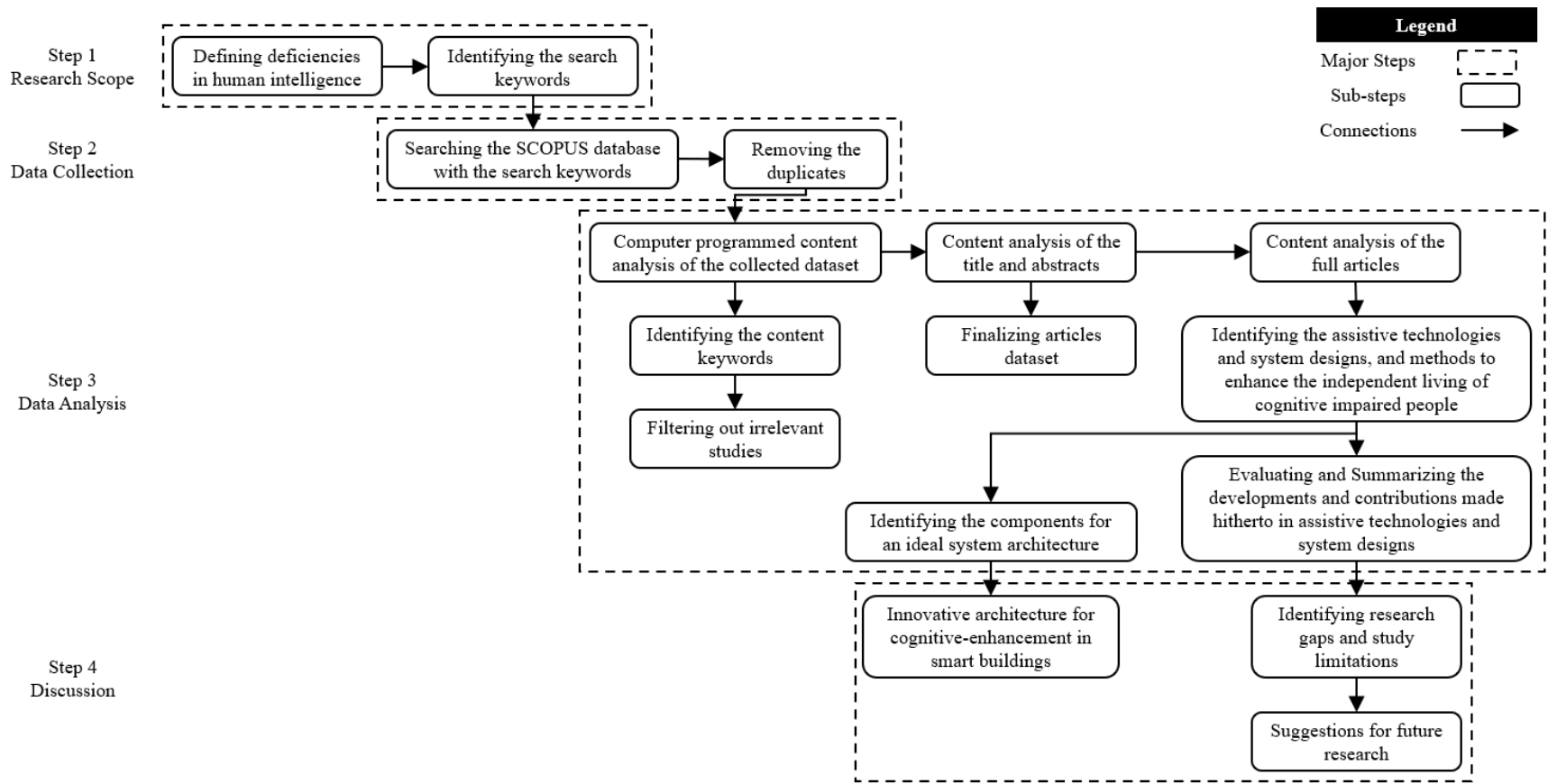

**Fig. 1 – Artile Search Method**

To collect articles related to the compensations and complements smart buildings offer to people with decreased intelligence, first, a general search was performed to identify the terms, which are called Search Keywords. The identified keywords were further used to perform the main search on academic databases. The authors only considered journal published articles and review papers to set a threshold for the quality of this study's materials for content analysis. Figure 1 depicts the research methods and the processes that the authors went through for this study.

**2.1 Research Scope**

Diminished intelligence is defined as any sort of medical condition influencing cognitive abilities which is called "Cognitive Disability" [20]. Other terminologies have been used for labeling and describing different causes of diminished intelligence, including cognitive impairment, cognitive disorder, and cognitive dysfunction. In this study, the general term "cognitive decline" is used so that it would be inclusive of all causes of cognitive disabilities, disorders, impairments, etc.

Smart homes and buildings enabled with IoT, or CPS can provide an alternative care option for those suffering from cognitive disabilities. Patients can remain at home and live independently of the caregivers with the help of assistive solutions embedded in the built environment [3]. In this research, the assistive solutions embedded within the smart buildings to increase the independency, and thus, QoL of the occupants suffering from cognitive disabilities are studied. The authors considered the studies from all around the world that only are published in English.

**2.2 Literature Search**

In the initial search, we permutated two groups of search terms with Boolean operator "OR" to search through the whole body of text in publication available on Scopus database (www.scopus.com). This database was specifically chosen since its comprehensiveness and the altogether access to academic publishers and online libraries such as Science Direct, Taylor & Francis, Wiley, PubMed, etc. The first group were defined to represent the cognitive disabilities and included "cognitive disability", "cognitive impairment", "cognitive disorder", and "cognitive dysfunction". The second group represented the scope of the environment and are "smart built environment", "intelligent built environment", "smart home", "intelligent building", "smart building", and "intelligent building". We identified 925 scientific articles which were written only in English, published from 2005 to 2020, and limited to technical and review papers published only in scientific journals to set a threshold for the quality of the search results.

To narrow down the result of the initial search to the articles related to the scope, the authors developed a Python script for text mining. An .xlsx file was downloaded from Scopus with information including author names, titles, published year, abstract, author keywords, and index keywords. First, articles with duplicated author names and titles were identified. 671 articles remained after removing the duplicates. To identify appropriate search terms, the authors combined the two keyword sections (author and index keywords). Abstracts were also preprocessed. The Authors initially stripped the abstracts from annotations and stop words including pronouns, transitions words, conjunctions, etc. Secondly, all the letters were changed into lowercase to create a consistent text data and avoid case-sensitive biases. Then, the most repeated words (Table 1) were calculated to extract the search terms related to the scope of the research. The search terms were used as an input for the Python code to search through the article's information. Similar to the initial search, the extracted keywords were divided into two main groups, one describing occupants' condition, and the other representing the type of built environment. The code identified 249 articles that had at least one combination of the two groups in either of their titles, abstract, author keywords, or index keywords. Any technical paper containing a combination of words from the Primary Words in the Table 1 with the corresponding words in the Descriptive Words column of the Table 1 were included in this step.

**Table 1**
The most repeated words to use in the search script

| Primary Words | Descriptive Words |
|---|---|
| Cognitive | Impairment, Decline, Disability, Dysfunction, Disorder, Defect, Deficit, Assistance |
| Disease | Alzheimer, Parkinson, Mental, Brain |
| Disorder | Neurocognitive, Mental, Memory |
| Dementia | Assessment, Mild, Syndrome |
| Disability | Intellectual, Neurological |
| Intelligent/ Intelligence | Building, Environment, Artificial, Ambient Sensor, Home, Assistance |
| Smart | Home, Environment, Hospital, Living, Monitoring, Habitat, Health, House |
| Living | Assisted, Independent, Assistive, Elderly, Sensorised |
| Activity | Recognition, Living, Human, Monitoring, Detection, Prediction |
| Behavior | Detection, Pattern |

> **Input 1** .xlsx file of initial search in academic databases with duplicates removed
> **Input 2** Column list of the file on which the script is going to search the extracted keywords
> **Input 3** The first set of search keywords
> **Input 4** The second set of search keywords
> **Output** .xlsx file of articles related to the scope of the research
> **Create** an empty local dataframe
> **For each** header in the column list
>     **For each** keyword#1 in the first set of keywords
>         **For each** keyword#2 in the second set of keywords
>             **Store** all the articles with a combination of keyword#1 and keyword#2 in the selected header in the dummy#2 dataframe
>             **Merge** the local and the dummy dataframes with duplicates removed

**Algorithm 1 – Pseudocode of the selection process**

## 2.3 Content Analysis

After reviewing the title and abstract of the 249 papers, 151 were in the scope of the research. In the next step, the text and content of the 151 identified articles were analyzed thoroughly. At last, 46 technical papers, 23 review papers, and 18 reports or survey studies remained which were fully in the scope of this study. Articles that only assessed the cognitive profile of the occupants without offering any assistive solutions or services, provided assistive solutions or services out of a living environment, did not address the cognitive profile of the occupants, or focus on other type of occupants e.g., caregivers instead of cognitively declined patients considered out of scope, and thus, were removed.

The authors reviewed each paper with regard to aspects: 1) outcome of the research, 2) the research method, 3) data collection method, 4) the algorithms that were utilized to process the data, 5) the demographic group that were studied, 6) requirements to live independently, 7) types of the assistive solutions and service that were offered, 8) how the solutions and services were delivered, 9) the research gaps that were covered by the studies, 10) the environment wherein the studies were conducted, , 11) whether any challenges, obstacles, or limitations were discussed, and 12) whether any direction for future research were suggested. Figure 2 presents the distribution of the reviewed technical papers on these aspects. Chart A illustrates the number of reviewed technical papers by their research and validation method. Chart B depicts where the studies were conducted; real environment is the representative of taking the whole system to patients' house, controlled environment is when researchers brought patients into a previously set up environment, and simulation is when the data is generated virtually or downloaded from existing data bases. Chart C represents the number of papers that proposed a system/framework/platform (as opposed to continue a previously conducted study), the discussed challenges and limitations the researchers faced during their work, and the suggestions for future research.

## 2.4 Originality

Utilization of technologies to assist the cognitively declined people has been evaluated and summarized before. Except one review paper, others had different scope including partial discussions on smart homes [7,14,21–25], evaluation of the usability, requirements, or issues related to the technology implementations [11,23,26–29], assessing the treatment responses to technologies [30], evaluation of technologies for a broader range of disabilities [5,31], evaluation of technologies for wandering [21,32–34], evaluation of other users such as caregivers [11,35], evaluation of specific function of smart buildings [12,36], partially discussed the services and solutions provided by the built environment [1]. The other literature review involves evaluating the developments in frameworks and sensor systems without addressing the use of CPS [8]. While cognitive impairments, disabilities, disorders, or dysfunctions can occur with different causes, the demographic groups of all the review papers are elder adults suffering from chronical disease such as Alzheimer or dementia.

    The remaining porting of the articles includes survey studies of the usability, implementation, or functionality of assistive technologies [15,37–40], reports or survey studies on the developments of the assistive technologies [4,41–47], reports on the ethical considerations of the technology implementations [48,49], and survey studies for need assessment of the end-users of assistive technologies [50,51].

    To our knowledge there is no comprehensive review that particularly aim to evaluate recent developments and contributions in the technological interventions that smart built environment can provide by adopting CPS, IoT-enabled sensing, and learning capabilities to compensate for the lack of intelligence in occupants. This research is more focused towards the assessing the utilization of CPS to offer a full independency of caregivers to occupants suffering from cognitive decline. A summary of the use of technologies to assist cognitively declined people is presented in Appendix I [52].

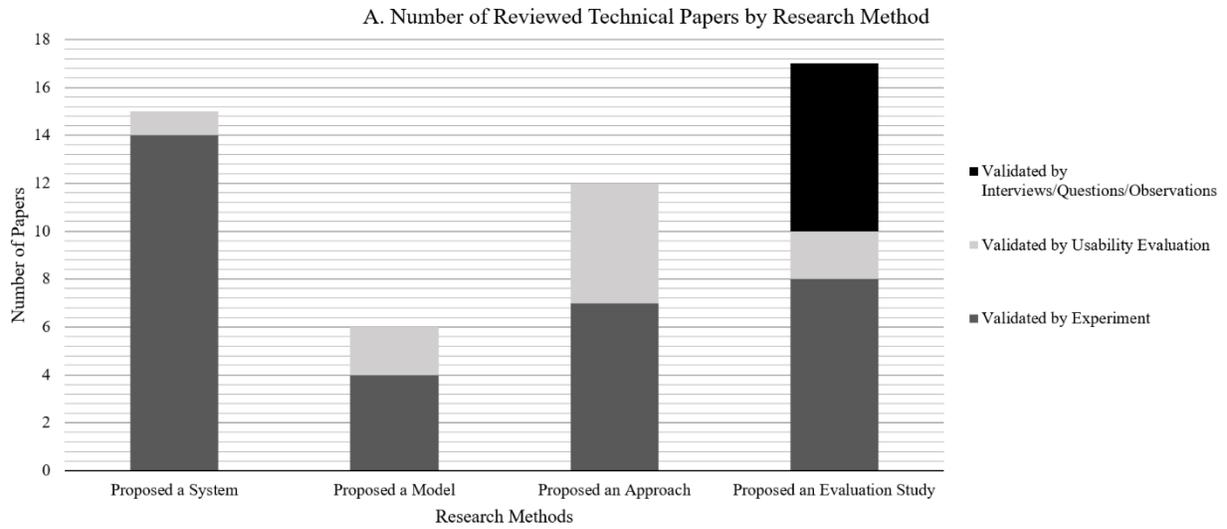
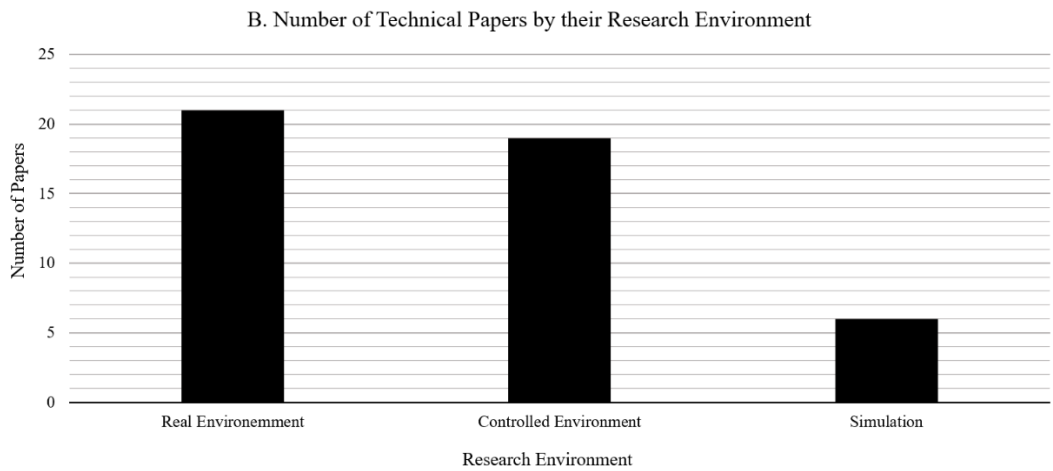
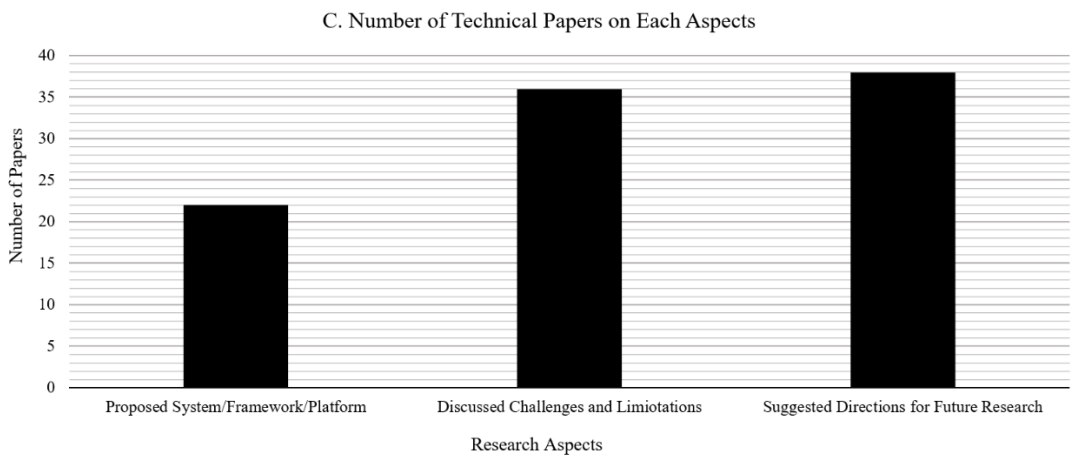

**Fig. 2 – Distribution of the reviewed technical papers on each aspect**

# 3 Overview of Existing Smart Building Technologies for Cognitively Declined Occupants

Smart buildings enabled with CPS and IoT can detect where the cognitively declined occupants lack intelligence [6,25,44]. The system processes the collected information and determines corresponding assistive services and solutions [53]. In the next step, the system controls deployed actuators to implement the determined assistive services and solutions to the physical environment, in form of reminders and technological interventions [16,18]. The assistive services and solutions that the smart buildings provide are intelligence compensations for either prevention or mitigation of unwanted situations helping those with cognitive decline to improve their quality of life (QoL) and simultaneously preserve their independence [4,27,54,55].In general, three components constitute the architecture of the intelligent systems in smart buildings: the *Data Collection Layer*, the *Processing Layer*, and the *Control Layer*. The first component, the *Data Collection Layer*, consists of different and various interconnected sensing technologies transmitting real-time information to the *Processing Layer*.

With the collected real-time data, the *Processing Layer* performs assessments and interpretations to derive meaningful solutions and decisions on how to provide assistive services and intelligence compensations. The solutions and services are normally identified based on three types of analysis that the *Processing Layer* carries out. They are 1) anomaly detection, 2) activity recognition, and 3) behavioral analysis. Anomaly detection is defined as detecting a condition or situation wherein the occupants are at risk of harm or injury e.g., when occupants wander, fall, etc. [56]. Activity recognition, or often times known as plan recognition, aims to comprehend occupants' actions and intentions through performing ADLs [57]. In behavioral monitoring, the *Processing Layer* evaluates the performance of the occupants performing various activities and identify patterns of behavior [58], or perform both the activity recognition and anomaly detection [6].

The identified solutions and services are then delivered to the occupants through either prompts or interventions. While the majority of the articles targeted a cognitively impaired demographic group in general [37,54,55,59–63], others carried out their studies specifically on a sub-category of cognitive impairment such as traumatic brain injury (TBI) [64,65], dementia [53,56,58,66–68], stroke rehabilitation [53], Alzheimer Disease [6,9], Parkinson Disease [69], Apraxia, and action disorganization syndrome [70].

The collected database of the technical and review papers is varyingly diverse in study domain, methodology, and application. Papers are from different domains including psychology, computer science, gerontology, electrical engineering, and even medicine. The most occurring journal is the "Journal of Ambient Intelligence and Humanized Computing" with four articles in the database. The methodologies range from a simple explanatory analysis of the collected sensing data to advanced algorithms and machine learning. The results of the evaluated methods are varying and, in some cases, contradicting. The focus and contribution of each paper, to some extent, are different than others. Most of the collected articles contributed to developing a thorough assistive system for cognitively declined occupants. However, their focuses range from introducing novel data collection methods, data processing techniques, or intervention approaches to evaluate the usability of proposed frameworks developed with existed methods, techniques, approaches. Nevertheless, the research findings regarding the identification and delivery methods of intelligent compensations are discussed in the following. Appendix II represents the summary of reviewed technical papers [52].

## 3.1 Anomaly Detection

A fall is defined when an individual come to rest on the ground or another lower level without any intentional movement, major intrinsic event or extrinsic force [71]. The processing layer recognizes falls when a threshold is exceeded while analyzing the real-time feed [60,72,73]. The thresholds are determined either empirically in a lab or during the learning phase with machine learning methods. Through utilization of a network of anchor points tagged subjects are located and their positions and movement data are collected. Thresholds are the average of the detected events in the previous 30-day period [72,73]. The employed algorithm detects anomalies based on the comparison of real-time feed with the determined thresholds [60].

Although it is complicated, more time-consuming, and requires extended resources, further studies should evaluate the use of probabilistic models and learning algorithms to predict falls. Abbate et al. [60] used Support Vector machines (SVM) to automatically monitor and predict events with a certain probability. Their system records brainwave data as well. The adopted learning technique identifies signal patterns under normal conditions. Then it tries to predict whether any future data point would differ from the pattern, which indicates a fall. On the other hand, Charlon et al. [72,73] analyzed their collected data with machine learning in two studies to detect falls without offering the prediction advantage. Fall detection is beneficial for the aged adults living alone. Caregivers or clinicians arrive after the system detects a fall and take care of the situation. However, the learning and prediction capabilities of the smart environment will even reduce the caregiving responsibilities further by preventing unwanted situations from happening. Moreover, Tchalla et al. [71] detected falls with an electronic bracelet. Because of the simplicity and not using any sort of learning techniques, such systems produce too many false positives.

Wandering is defined as a locomotion behavior that is repetitive with temporally disoriented nature accompanied by lapping, random, or pacing patterns [17]. It happens either in sleep or as a result of stress [47,56]. At night, when the pressure sensors under the mattress cannot detect the occupant and motion sensors on the bedroom ceiling or doorway are triggered, the system interprets that the occupant is wandering [38,59,74,75]. Smart buildings can learn the circadian rhythm of the occupants to detect nighttime wandering [38]. Given the pre-defined thresholds based on occupants' cognitive profile or clinicians' expert opinion, smart buildings can detect high stress levels and sleep-related abnormalities such as sleep interruptions, latencies, and duration [56]. Falls and wandering can also be recognized with wearable sensors such as electronic bracelets [71], or a combination of wearable tags and anchor nodes in a radio localization system [76].

Wandering has been detected using a network of sensors in most cases [59,71,74,75]. However, by applying learning techniques, the system can monitor multiple occupants and can estimate movement trajectories. Kolakowski et al. [76] estimated the trajectory of the occupants' wandering indoors using an extended Kalman filter based algorithm with high accuracy. The trajectory estimation prevents any wandering related incidents. More advanced learning techniques prevent false positives with adequate training data.

### 3.2 Activity Recognition

Similar to the categorization carried out by [2], the approaches that have been taken to conduct activity recognition fall into three categories: 1) deterministic, 2) non-learning probabilistic, and 3) learning approaches.

Deterministic approach is observing occupants' actions based on a pre-defined activity pattern. Each activity consists of several tasks required to be performed in specific orders to complete the activity successfully. Therefore, an activity pattern is a sequence of various tasks or steps, which, if done in order, will result in the successful completion of the corresponding activity. The *Processing Layer* compares real-time feed of on-going activity with the pre-defined pattern to detect deviations [55,65]. Belley et al. [55] proposed an algorithmic approach that measures the electric current and voltage of the home appliances to detect incorrect usage of appliances for preparing breakfast. Pinard et al. [65] analyzed the real-time data of the occupants via a context-aware approach to ensure they are performing the activities as instructed and to recognize abnormalities.

Non-learning probabilistic approach aims to predict the following tasks by observing few initial ones with a pre-defined pattern. An activity can be recognized through the activation of a particular order of sensors that are used for monitoring relevant movements. When sensors detect initial steps of an activity or partially collect data, the *Processing Layer* calculates the probability of possible activity patterns to predict the following steps [2,3,63,74,77]. The classic probabilistic method, Markov Decision-making Process (MDP)-based methods, is a commonly used approach to calculating the probability of the upcoming steps [53,66,70].

Mihailidis et al. [66] employed a tracking system that utilizes a Bayesian sequential estimation technique to process the images captured by video cameras for detecting activity-specific gestures when washing hands. Then, A Partially Observable Markov Decision Process (POMDP) was applied to obtain user-specific intelligent inferences. Hoey et al. [53] developed a unifying model based on POMDP . They modeled the tasks as the elements of a user's activities that are considered as reactions to a caregiver's actions. Pastorino et al. [70] identified the occupants' activities by an Action Recognition (AR) algorithm during activities' naturalistic execution. After an activity is identified, Task Model (TM) algorithm will be employed to detect possible mistakes committed by a subject. Additionally, TM predicted the next task to be performed for the subject to follow the action goal. To decode continuous signals of the sensors, Pastorino et al. (2014) adopted Markov Decision Process (MDP), upon which the TM verifies the tasks recognized by the AR and follow the subject's progress in an activity.

Bouchard et al. [2] recognized plausible abnormalities in performing ADLs by using lattice theory and an action model based on description logics that turns the activity recognition into classification. Roy et al. [57] used possibility theory for activity recognition and assisting occupants in their ADLs. Their proposed model recognizes the sequence of tasks performed to complete an activity. Tasks are formed based on a context-transition model that quantifies the transitions between contexts, obtained by task realization, with a possibility value. Aloulou et al. [74] developed a next-hour-activity-prediction service which predicts the patients activity using Decision Tree algorithms. Chaurasia et al. [3] incorporated a duration-based probabilistic model into their proposed system that assists the occupants in making drinks. The occupant's activity pattern is described with the joint probability distributions of the tasks within the pattern. The sequence of tasks completed, and the time spent on the activity were used to calculate the distributions using maximum likelihood estimation. Through combination of RFID and the electrical load signature of the appliances, Fortin-Simard et al. [77] identified tasks of an activity with the spatial interaction of the occupant with objects or appliances. Then, plausible activities associated with likelihoods were drawn with the use of Bayesian recognition process.

With the help of sensing technologies and machine learning algorithms, smart building can learn patterns occupants normally follow to perform a specific task forming a learning approach. The systems can then detect that the occupants perform an activity erroneously as a result of deviating from the learnt pattern [62,63]. To create activity models, Moutacalli et al. [63] collected ordered activated sensors with duration time between adjacent sensors. Activated adjacent sensors represent occupants' actions. The frequent sensor orders were clustered into activities using Fuzzy C-Means. If an out-of-order sensor were activated, the system would recognize that the occupant made a mistake. Hao et al. [62] proposed an inference engine which learns the sequential tasks of an activity to form patterns. The engine transformed the activity prediction into a graph searching problem solving with a new half-duplex graph searching algorithm (HDGS) based on Breadth-first search (BFS). Predictions are inaccurate at the initial stages of executing an activity, and some activities have semantically similar tasks and sequences. Therefore, Hao et al. [62] applied approximate predictions through clustering the tasks of interests according to their semantic similarities using Formal concept analysis (FCA).

To thoroughly increase the QoL for cognitively declined occupants, smart buildings must possess learning and predictive capabilities. Since occupants' activities are stochastic in nature, even a simple routine can be executed in different ways causing the deterministic approaches to fail to detect ADLs correctly. A combination of learning and predictive capabilities can consider the stochastic nature and adapt the decision-making process of the systems to changes in the execution of ADLs. Moreover, Predictive capabilities better prepare the *Processing Layer* to perform more robust decision-making. The systems will be a step ahead in understanding occupants' behavior. Therefore, more accurate services and solutions can be offered.

### 3.3 Behavioral Monitoring

A sequence of tasks forms an activity, and activities construct a behavior. To recognize behavioral patterns and monitor the occupant, the *Processing Layer* either compares the real-time feed with pre-defined sets of patterns or rules (deterministic), predicts the following activities in a behavioral pattern (non-learning probabilistic), or learn the pattern of the occupants' behaviors (learning).

The classic MDP-based algorithms are the probabilistic approach that Najjar et al. [78,79] and Chu et al. [80] employed to identify and recognize multiple activities of occupants Najjar et al. [78,79] equipped their recognition module with reinforced learning. To resolve the ambiguous sensor data issue, Chu et al. [80] used POMDP and proposed a non-learning heuristic approach based on a dual control algorithm using selective-inquiry to solve the POMDP. Although the use of MDP as probabilistic method is advantageous, the method is computationally exhaustive and demanding. Lam et al. [6] proposed a learning activity monitoring system with the help of classification algorithms including SVM, Random Forests (RF), and Naive Bayes (NB) along with a context-based approach to improve the accuracy of the activity detection. Among the algorithms, SVM showed the highest accuracy. The use of machine learning algorithms must be evaluated more to find an alternative to MDP that possesses both the predictive and learning capabilities.

Stucki et al. [9] presented a web-based system to identify and classify the activities of daily living (ADL) based on a rule-based forward chaining inference engine. The collected data were compared to a set of predefined rules to evaluate whether the data fit into a corresponding ADL. Lazarou et al. [54,61] and Stavropoulos et al. [68] used semantic interpretation and fusion to extract meaningful behavioral patterns and complex activities from raw data. They deployed processing methods ranging from simple retrieval of data to complex activity recognition and computer vision algorithms introduced by previous studies. The system takes advantage of a set of predefined rules (expressed in SPARQL) which clinicians adjusted based on each occupant's profile. To detect a daily routine, Enshaeifar et al. [67] applied pattern recognition algorithms including ruled-based reasoning algorithms and adaptive learning. These studies introduced a deterministic approach towards monitoring ADLs. Each of which is capable of monitoring multiple activities with high accuracy. However, in time, execution of ADLs goes through slight to moderate changes if not implemented and planned correctly in the system, errors would occur. This can be avoided with the help of learning techniques instead of rule- or logic-based approaches.

Another composition of behavior is anomalies. Lotfi et al. [81] transformed the anomaly detection into a classification problem and evaluated a solution based on different clustering algorithms, including self-organizing maps (SOM), K-means clustering, and fuzzy C-means (FCM). Charlon et al. [72] initially extracted a behavioral model from movement and activity data with the help of supervised classification-based learning techniques. Danger detection thresholds were then determined based on the average of detected events over 30 days. Real-time events were evaluated, and anomalies were recognized based on the thresholds. Alvarez et al. [69] developed a system that utilized movement trajectories with a Sparse Autoencoder (SAE) algorithm to differentiate anomalies such as wandering from normal behaviors. To consider the reliability of the data collected from low-level subsystems, they applied a weighted probabilistic model. In a case that sufficient training data is available, a Bayesian Network (BN) was applied to retrieve high-level information such as executed tasks of an activity and their temporal data. Moreover, they employed Recurrent Neural Networks (RNN) to detect the Freeze of Gait (FoG). Keum et al. [58] proposed a system that constructed behavior patterns based on analyzing the sequential timing of single activities to recognize abnormalities in ADLs using edge computing. Enshaeifar et al. [82] developed two algorithms to detect Urinary Tract Infections (UTI) and changes in activity patterns of the occupants to provide personalized assistive services using a Non-negative Matrix Factorization (NMF) and an Isolation Forest (iForest) technique.

To offer full independence from the caregivers, smart buildings must be able to monitor various aspects of the occupants' lives or in other words monitor the occupants' behavior. Occupant behavior is a combination of simple activities such as washing hands, complex ADLs such as cooking, and anomaly detection to prevent the occurrence of events such as fall and wandering-related incidents. Smart buildings equipped with advanced *Data Collection* and *Processing* layers capable of behavioral modeling with learning and predictive techniques will take over the majority of caregiving responsibilities.

## 3.4 Prompting

In the context of smart buildings, a prompt is an indication, a reminder, or a suggestion delivered to a user based on time, context, or acquired intelligence assisting the user with successfully completing an on-going activity [4,55,83]. Smart buildings normally offer four types of prompting: audio, visual, audiovisual, and lighting [55]. The first is verbal cues, reminders, or step-by-step instructions delivered through planted speakers [64,84]. The second offers non-verbal or pictorial assists to the occupants via screen of smartphones, TVs, tablets, etc. [6,67,69,79]. Audiovisual prompting is the combination of both [77,85–87]. Prompting through lights involve using a laser or a bulb that changes colors to indicate the status of the situation [55,79], flashes or points to help the occupant locate an object [2,55,76]. Prompting is used in terms of alert systems as well. In cases when the *Processing Layer* detects anomalies, the *Control Layer* is signaled to send emergency alerts and messages to caregivers, nurses, or hospitals [60,76,88].

Prompts must be chosen according to the occupants cognitive profile and the mistakes they make in performing ADLs [55]. Fortin-Simard et al. [77] and Belley et al. [55] categorized mistakes into omission, sequence, perseveration, temporal, and cognitive overload. Omission happens when the user forgets to execute a task or step in an activity. Sequence is the disruption of the execution order of tasks to complete an activity. Perseveration occurs as the occupant persists on a same task or step. Temporal is the inconsistencies with the predefined time constraints of performing tasks and completing activities. Lastly, cognitive overload happens when the user tries to perform multitasking which causes distractions and errors due to cognitive decline. The *Processing Layer*, first, identifies the type of committed mistake, and then, signals the proper instructions to the *Control Layer* to deliver through prompting.

Due to the fact that memory recedes in cognitively declined occupants, they struggle with short-term memory loss, scheduling activities, and learning from their past mistakes [4,6,62]. Moreover, the forgetfulness often results in misplacing the personal belongings, disrupting the necessary medication, and poor nutrition [6,61,76]. In such situations, the *Processing Layer* signals proper reminders to the *Control Layer* to notify the occupant the previously made mistakes, their plans, their medication, and assist them in locating a mislaid object.

### 3.5 Intervention

Devices such as actuators and effectors have been deployed to intervene the built environment to keep occupants safe [38,47,71,88], or guide them through completion of an activity [86]. Gerka et al. [47,88] deployed a stove deactivation system to prevent fire hazards. The system monitors the stove with motion detectors and temperature sensors. If the temperature exceeds a threshold and the motion sensors did not detect any activities, it automatically turns the stove off. Tchalla et al. [71] and Martin et al. [38] planted actuators to provide lighting guidance in cases of nighttime wandering to prevent falls. The motion detectors planted in the bedroom, or the pressure sensors planted under the mattress detects movement, and then actuators are triggered to light the path to bathrooms or other rooms of the house. The system that Martin et al. (2013) deployed provides automatic therapeutic interventions through playing music when occupants were having sleeping troubles to prevent wandering caused by stress. Ficocelli and Nejat [86] designed and developed an initial prototype of an interactive assistive system that helps the occupants with performing kitchen-based ADLs. The system features an automated cabinet system that helps storing and retrieving kitchenware and ingredients of meal preparation during cooking. The cabinets can move towards the occupants to provide the access to the items they want or to help them store the items in the right places. The system initiates and receives requests/orders with direct verbal prompts of the user. The speech recognition module incorporates Hidden Markov Models to recognize the user's prompts.

## 4 Research gaps and recommendations for future research

Smart buildings have the potential in providing an autonomy for the cognitively declined occupants to decrease their dependency on caregivers. However, many important issues have not been addressed, and more robust systems are yet to be invented, designed, and developed. This section discusses the identified challenges, gaps, limitations, and the recommendations provided for future research in four domains: acceptability and usability of the smart building systems, ethical concerns of the users towards using smart building systems, comprehensiveness of the systems, and system evaluation and human-in-the-loop.

## 4.1 Usability and Acceptability

Challenges such as acceptability, durability, ease of use, or power requirements are normally accompanied with the introduction and utilization of new technologies [1]. Smart homes, as an example of such technologies, are often complex and require specific skills to install and operate [9]. Many complex and advanced systems have been only tested and implemented in a controlled environment where trained experts were available during experiments to operate and maintain the systems. These conditions are far from patients' living environment and are not sufficient for acceptability and usability evaluation. End-users have reported technical issues such as false alarms, wrong prompts, poor connectivity, low batteries, high costs, etc. each of which affects the usability of the systems [14,21,89]. Successful implementation of smart homes necessitates organized, structured, and individually accustomed procedures. Regular follow-ups are required to provide the necessary support [15].

With cognitively declined people as the end-user, the acceptability of such system becomes even more important. One major barrier preventing the acceptance of assistive systems in smart built environments is the lack of familiarity that the end-users have with technologies in general [11,54,70]. The systematic design of smart homes significantly impacts the difficulties the end-users face to learn how to use the systems [15]. Designs must be simplistic and easy-to-use with little involvement of the end-users in system's installation, operation, and maintenance [9,21,38]. Holthe et al. [26] defined usability of any technology as "user friendliness, usefulness, and effectiveness, and by the extent to which a product can help a user to achieve a specific goal." One approach to increase the acceptability of such technologies is to incorporate the principles of "universal design" which is a broad concept in the design spectrum of any product or type of environment. With a universal design, cognitively declined people are under the impression that the implemented system has been used with many users resulting an incline towards accepting the technology.

## 4.2 Ethical concerns

Occupants reported that they avoid using smart home systems for ethical issues like privacy concerns and intrusiveness. To monitor the occupants, smart homes require the utilization of video- and audio-based sensors. These systems are called intrusive [25,55,69], invasive [13,53], or obtrusive [4,31] throughout the literature. The use of cameras or microphone to collect behavioral data necessitates end-users' consent, which is not often possible with the cognitively declined patients. In some cases, the researchers ignored obtaining the consent and assumed the end-users will accept the original design of the introduced system with vision- and audio-based sensing technologies. Survey studies show that users are aware of the benefits of the new technologies and accept those technologies easily if proved unintrusive or their consent is sought.

Using technologies that are known as surveillance technologies for controlling people implies that cognitively declined people are required to be controlled and restrained. The use of IoT, wireless devices, and information storage facilities, always introduce potential privacy, data ownership, and cybersecurity issues. Data ownership and security should be first clarified and structured before implementation of the smart home technologies. Although, there has not been a comprehensive and effective approach to address these gaps, the following criteria should be considered in the design of such technologies: 1) cognitively declined people's ability of to deliver informed consent, 2) how to protect their privacy, 3) what is the confidentiality level of the collected data, 4) who can have access to the data, and who owns it.

### 4.3 Comprehensiveness of the System

Many developed assistive systems are use-case-oriented, and a set of services and scenarios are predefined to respond to specific users' needs. These systems cannot evolve when new requirements emerge, or a new user is introduced to the environment. In addition, some assistive technologies only evaluate single needs of the occupants instead of providing services and solutions for multiple requirements that occupants have during various stages of their condition, such as wandering, falls, sleep quality, and daily tasks. There is variability among different individuals with different cognitive profiles. Comprehensive cognitive assessments should be incorporated into the systems' implementation so that classification of cognitive profiles can be performed, and more personalized assistive services can be delivered to meet the users' preferences.

Most of the implemented smart home systems are not capable of distinguish between the user and other occupants. Multiple occupancy recognition has been reported an issue in many developed systems. This issue should be addressed without disturbing the users' sense of control. More research is required so that the systems can recognize activation of several sensors and remain focused on the monitored subject.

In many reviewed studies, it was assumed that smart home systems have been developed based on the incorrect assumption that the activity and behavioral pattern of the occupants will remain consistent in time. Therefore, only one or several specific patterns have been developed for each monitored activity or behavior. To enable the smart homes with learning capabilities and mechanisms, probabilistic and learning analytical methods, such as combining deep learning models with probabilistic machine learning techniques, should be evaluated and incorporated into the systems. In spite of few applications of machine learning and probabilistic methods, uncertainty, imprecision, and fuzziness related to the collected information have not been considered properly. Developing predictive and incremental learning models also provide more insights and information from monitoring physiological data [90].

## 4.4 System evaluation and Human-in-the-Loop

Assistive technologies and smart homes provide services to cognitively declined people as a substitute for caregivers. Accordingly, to prove their adequacy, maturity, and efficacy, clinical trials are necessary. One drawback of the developed smart home systems is lack of clinical expert feedback and proof during the experimental phase of the systems. In some cases, developers launched their prototype without a proper testing trial [8,57,85]. Linguistic and neuropsychological research is a must especially for the design and effective delivery of prompts and interventions. Proper testing trials must include laboratory testing with the presence of clinicians. Since conducting experiments in a controlled environment such as a laboratory eliminates variability of the real environment and produces only polished data, experiments in patients' daily living environment in need. Several reviewed studies that involved experiments only tested their developed systems in a controlled testbed, resulting in a questionable efficiency of the systems in a real environment such as the patients' houses.

Conducting experiments in a controlled environment also restrains the number of participants. Many reported that the number of participants was not sufficient to present conclusive evidence [14,77,88]. Finding subjects to construct a reliable population size was reported complicated and challenging. A small population size only leads to general and in some cases derivative findings. Another challenge is that small samples do not represent the whole population efficiently. Variability is sacrificed as a result of a homogenous sample population. More studies are necessary with large sample sizes that include different cognitive profiles and various environments.

When developing the assistive technologies, the concept and approach of user-centered design should be adopted more. The users' perspectives were not often considered in the reviewed studies. The first-person account is necessary in introducing and development of new technologies. System designers and developers should consider and increase the role of the end-users in the whole process. Evaluation studies of users' needs are not sufficient if individuals are not directly included in the design and development process. As Czarnuch et al. [91] stated, the end-users should be involved in all the stages of design, evaluation, and deployment of the assistive technologies including: needs assessment; idea generation; device prototyping; and efficacy testing. In experiments, subjects who attempted to communicate with the testing systems but did not receive interactive responses tend to have negative opinions on the proposed systems. On the other hand, subjects presented positive body language and facial expressions when the experimenters intervened to communicate with the subjects. Smart homes with prompting as their delivery method should be equipped with interactive user interfaces so that appropriate responses would be generated in any bidirectional communication attempted.

The content of the delivered prompts should be tailored to the users' cognitive profile to increase positive emotions and encourage task completion. There is a substantial need to study prompts and evaluate the best phrasing of prompts, whether questions should be asks, orders should be given, or suggestions should be offered, what the optimal number of words are, and what the best users' responses are.

**4.5 Implementation of Cyber-Physical Systems**

Building enhanced with IoT-enabled sensors, actuators, and effectors are smart buildings sharing features of a Cyber-Physical System (CPS) [18]. Such buildings can measure the physical built environment, collect the occupants' related measurements, and transmit the data to processors to further control the built environment functionality. Machine learning tools are normally adopted to enhance computational performance. Such integration of physical devices with cyber components (computation and communication capabilities) forms an intelligent analytical system responsive to the dynamic changes in the real world and occupants' behavior [18,92]. With the help of actuators and effectors, the physical built environment can respond to the occupants' needs, and thus, add more intelligence to their lives [93].

In the context of smart buildings, CPS has a wide variety of applications, such as smart medical technology, ambient assisted living, indoor environmental control, and energy management [18,93]. Ambient assisted living (AAL) environment is a combination of assistive technologies, solutions, and services that positively influence the occupants' QoL [31]. Cognitively declined occupants are dependent on others (caregivers) to carry out their ADLs. CPS-enabled smart buildings are capable of complementing their lack of intelligence and increase their independence, and thus, their autonomy in performing everyday activities. Such buildings can monitor the occupants and provide assistive services and solutions through prompting or intervention.

Prompting has been more effective when the severity of the cognitive decline is at a lower level. With the severity of the disease increasing, higher-level prompting is used. However, that is only effective to a specific point. Patients with a higher severity of the illness require a greater number of prompts which makes them agitated. Numerous prompting occasionally results in a failure in completing tasks [94]. At this point, interventions are required. Without utilization of any sort of technologies, caregivers are responsible to assist the cognitively declined people with their ADLs.

To increase the autonomy of the occupants, few limited technologies has been introduced with the ability of replacing the caregivers' responsibilities in specific activities. A range of robotic systems has also been introduced which play the companion role for the occupants [14].

Studies discussed have been introduced a form of CPS integration in the smart building concept. Technological interventions developed range from simple verbal cues delivered through planted speakers [74] to the use of embedded robotics in moving cabinets to the occupants [86]. However, the application of CPS in smart home domain has not been studied and researched appropriately.

# 5 A Conceptual Framework for CPS-enabled Smart Homes

In this section, the authors propose a new conceptual framework to structure an ideal system architecture for enabling smart homes with CPS within the context of AALs in smart residential buildings. The framework, inspired by the existing developments, can serve as a guide for future research. As presented in Fig. 3, the framework consists of three main components: *Knowledge Base*, *Processing Layer*, and *Control Layer*.

The *Knowledge Base* is composed of the IoT-enabled sensing that collects, transmits, and stores real-time data. The sensing technologies used hitherto can be classified into two major groups: Wearable and Non-Wearable. Smart bracelets, smart watches, or tags that can be worn by the end-users are considered wearable sensors. Researchers have used such sensors to localize, detect the movements, and collect the physiological signs and parameters of the subjects. With the consent of the end-users, deploying a combination of both wearable and non-wearable sensors provides sufficient behavioral data for the *Processing Layer* to learn the behavioral patterns and produce tailored assistive solutions and services. Plug&Play mechanism enables attachment of new sensors without reconfiguration of the whole system [59]. The mechanism can discover and connect newly deployed sensors to the existing network at runtime. The *Knowledge Base* also includes input/output (I/O) devices to provide the human-in-the-loop capabilities for the systems to interact with the end-users via voice commands, buttons, or other forms of input. The I/O devices can receive end-users' and experts' feedback to further improve or ameliorate the solutions and services generated by the system. The feedback is received through specific interfaces developed separately for occupants, caregivers, and clinicians.

Probabilistic ML and AI algorithms and tools are used in the *Processing Layer* to enable learning and predictive capabilities. Learning capabilities are essential to remove the use of rule/agent/logic-based algorithms. Although beneficial, they may fail to consider changes in behavior patterns. They are set to respond to specific users' needs, and thus, are use-case-oriented. Learning capability can learn the behavioral pattern of the user to recognize different activities, address various needs, and detect anomalies. It can also identify multiple occupants to distinguish between the users and other occupants or offer assistive services in cases of multiple occupants. Moreover, assessments of the occupants' cognitive profiles in different stages of their condition can be extracted from the real-time data. Predictive capability offers the opportunity for the system to be always prepared one step ahead. Therefore, the *Processing Layer* can predict the following tasks of a recognized activity and guide the user through completion. Also, the system can prevent unwanted events such as falling from happening. Markov Chain Process (MCP) or Markov Decision Process (MDP) and its derivatives such as Hidden Markov Model (HMM) are proper examples of a probabilistic model that is also used in Reinforcement Learning.

Despite the intelligence shortcomings of the occupants, occasionally they are in complete control of the modifications they make in performing ADLs. Such cases hinder the activity prediction from acquiring high accuracy [55,64]. To resolve the challenge, I/O devices provide the occupants with triggering options so that they can self-initiate specific assistive services such as prompting. Reminders of such options should be set to take out the possibility of forgetfulness. Consequently, more tailored assistive solutions and services are identified for the users when those capabilities are enabled. The content of the prompts is personalized to users' cognitive profiles, and the services can be adapted and updated when changes are recorded in the behavioral patterns of the users.

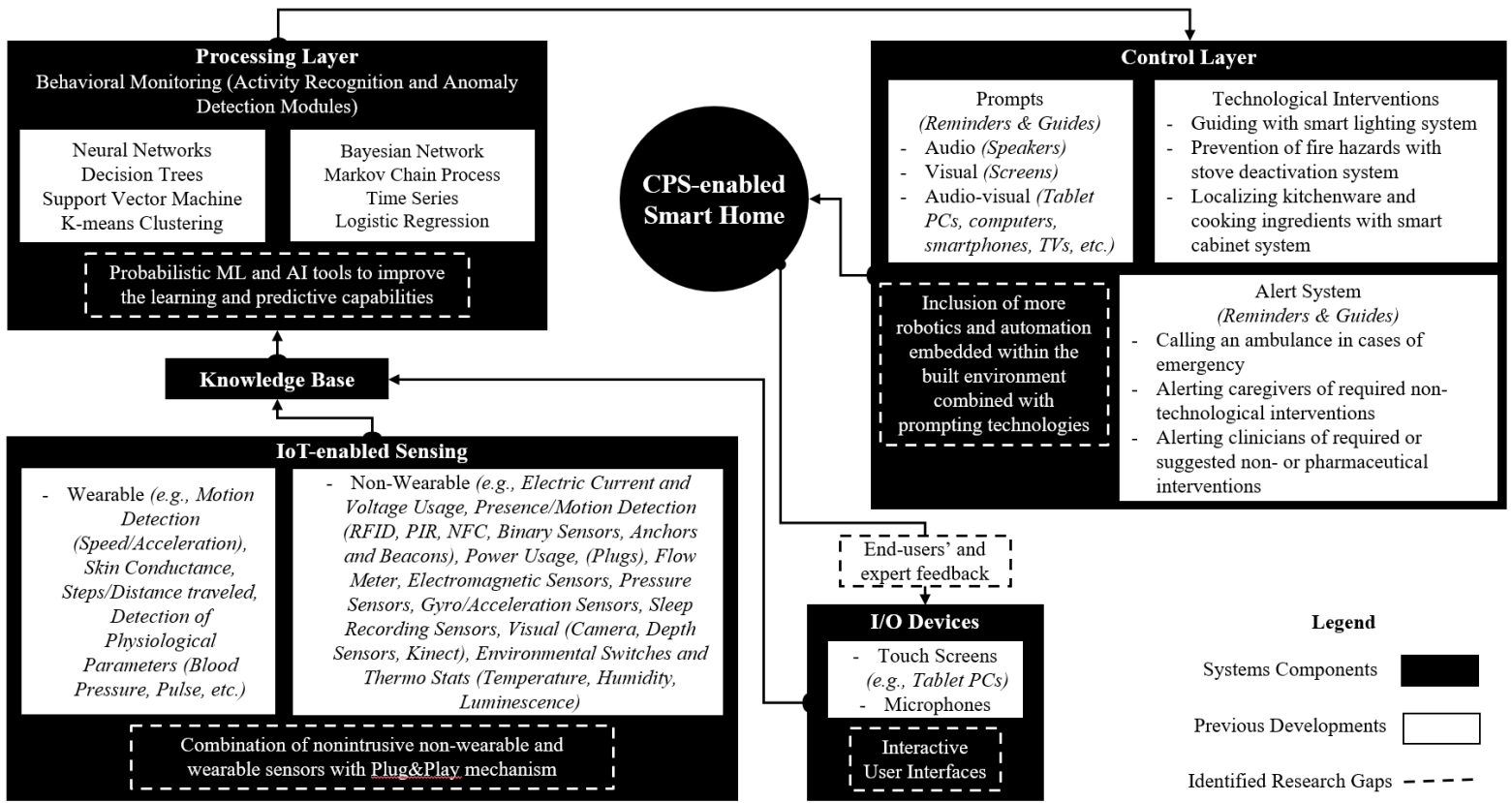

Fig. 3 – Conceptual framework for CPS-enabled smart homes

CPS involves problem solving and control of a physical environment [16]. CPS improves the performance of an IoT system via connecting the sensing technology with processors and actuators to allow carrying out actions in, interacting with, and even changing the physical world [18]. This improvement is necessary to achieve an increase in the independence and quality of life for cognitively declined occupants without relying on caregivers [17,61,70]. The *Control Layer* is where the Physical Twin of a CPS is implemented. When solely prompting is not effective, the smart built environment should be capable of offering different forms of technological interventions with the help of embedded robotic and automation. Positioning the misplaced items and wandering prevention through interactive verbal communication with the system [76,86], automatic cabinet system, appliance activation/deactivation, cooking guide, and planner systems in the kitchen [47,65,77,86], automated handwashing guide [66], and automated lighting system for pathfinding when nighttime wandering occurs [72] are the examples of technological interventions developed previously. In cases of emergency when interventions by caregivers or clinicians are required, they should be alerted immediately by the system [72]. Future studies can focus their efforts to evaluate the incorporation of more embedded robotics and automation such as altering the physical built environment and the floor layout to prevent fatal accidents when falling.

**Conclusion**

This research aims to summarize and evaluate the existing research studies on the ubiquitous intelligence smart buildings can offer to the cognitively declined occupants. The studies evaluated in this research have indicated that the developed ambient intelligent systems have capabilities in assisting occupants with cognitive disabilities in various ways to increase their independence and quality of life. A typical occupant-assisting intelligent system consists of three main layers: the *Data Collection Layer*, the *Processing Layer*, and the *Control Layer*. The systems collect and transmit data through IoT-enabled sensing, process and analyze the data, provide information to the caregivers and the patients, and automate some caregiving tasks.

The authors identified five major research gaps. User-centered designs and interfaces are needed to improve the system acceptability and usability. The ethical concerns regarding the sensing technologies should be addressed to ensure the privacy of the occupants. Since the end-users are cognitively declined, obtaining their consent is not always possible. To derive more complex and comprehensive assistive services, smart building applications need the capability to evolve with the emergence of new requirements and address multiple needs of the occupants. Multiple occupancy is another gap of the existing body of research.

Previous studies have used machine learning and AI for anomaly detection, activity recognition, and behavioral monitoring to identify appropriate assistive services and solutions. However, there are needs to increase the incorporation of the predictive methods and algorithms to provide the *Processing Layer* with both learning and predictive capabilities. Various testing phases are required in both controlled environment, such as laboratories, and real environments, such as patients' homes, to integrate the user's and expert's perceptions of the designed systems. In addition to a universal system design, the final product, which is the assistive services and solutions, must be tailored and personalized based on occupants' cognitive profiles and assessments. Most of the reviewed studies contributed their effort into only delivering the assistive services and solutions through prompting. Despite the valuable contributions, prompting is only proved to be effective to certain cognitive profiles. With a more advance cognitive decline, prompts may not be helpful and even make the user agitated. There has been a gap in developing sufficient *Control Layer* that can offer technological interventions. To fully support the cognitively declined occupants in performing ADLs and automate the caregiving responsibilities, more focus should be put into in-depth analysis and development of a comprehensive *Control Layer* for CPS-enabled smart buildings.